\documentclass[aps,floatfix,prd,showpacs]{revtex4}
\usepackage{graphicx}
\usepackage{dcolumn}
\usepackage{bm}

\begin{document}

\def\be{\begin{equation}}
\def\ee{\end{equation}}

\title{QED3 at Finite Temperature and Density}
\author{Pok Man Lo}
\affiliation{
GSI Helmholtzzentrum f\"{u}r Schwerionenforschung GmbH, Planckstra\ss{e} 1, 64291 Darmstadt, Germany}
\author{Eric S. Swanson}
\affiliation{
Department of Physics and Astronomy, 
University of Pittsburgh, 
Pittsburgh, PA 15260, 
USA.}

\date{\today}

\begin{abstract}
Schwinger-Dyson equations are used to study the phase diagram of QED in three dimensions. This computation is made with full frequency-dependence in the two-point function gap equations for the first time. We also demonstrate that reliable results are attainable in spite of an infrared divergence that is endemic to the theory. A theoretically sound method for dealing with cutoff ultraviolet regulators is presented. Finally, it is shown that the quenched and instantaneous approximations often used in the literature are inaccurate.
\end{abstract}

\maketitle

\section{Introduction}

A number of novel features of QED in three dimensions (QED3) has attracted attention to this theory.
For example, high temperature QCD can be represented as the dimensionally reduced QCD3. If the number of quark flavours is large, the nonabelian behaviour of the theory is suppressed and it may be approximated as QED3\cite{pis}. Massless QED3 in the large $N_f$ limit generates dynamical fermion masses that are suppressed exponentially in the fermion number. Thus this theory illustrates how large mass hierarchies can be dynamically generated\cite{Appelquist}, which is of interest in the construction of model field theories.

More recently, QED3 has been used as a model for three dimensional condensed matter systems. Examples include applications to high $T_c$ superconductors, where the relevant
dynamics is thought to be isolated to copper-oxygen planes in the cuprate\cite{highTc}. It is also considered as a gauge formulation of the 2+1 dimensional Heisenberg spin model\cite{richert} and a possible model for graphene\cite{graph} or spin-ice\cite{spin-ice}.

It is possible to introduce a topological Chern-Simons-like photon mass term to the theory in three dimensions\cite{originals}. This term breaks parity and time reversal symmetries. 
A nonzero photon mass induces a finite fermion mass at one-loop (and vice versa)\cite{DJT,Appelquist-parity}. This raises the interesting possibility that parity symmetry can be  spontaneously broken in the massless theory. This question was first examined by Appelquist {\it et al.} many years ago\cite{Appelquist-parity}. They concluded that a finite fermion mass was dynamically generated, but that these masses appear in pairs of opposite sign, thereby maintaining the parity symmetry of the vacuum and a massless photon. This issue was recently re-examined with numerical solutions to the Schwinger-Dyson equations, which demonstrated a surprising parity-breaking solution\cite{LS1}.

The application of QED3 to problems in condensed matter naturally raises the issue of determining its properties at finite temperature and density.  Here, we will focus on computing the dynamical fermion mass, photon self-energy, and the phase diagram for spontaneous chiral symmetry breaking.

Unfortunately, solving Schwinger-Dyson equations (SDE) at finite temperature (or density) represents a formidable technical problem, being equivalent to solving tens or hundreds of coupled zero-temperature SDEs. Thus, previous work in this area has employed a number simplifying assumptions. Chief among these is the instantaneous approximation, where frequency-dependence in the fermionic interaction is neglected\cite{D00}. This permits exact evaluation of the temperature-dependence and a subsequent reduction of the numerical task by an order (or two) of magnitude. Since no small parameter underpins the accuracy of the instantaneous approximation, it must be justified {\it a posteriori}. We make this comparison here for the first time, and find that the instantaneous approximation is not reliable.

Another technical issue bedevils the study of QED3 at finite temperature; namely, an infrared divergence appears in the SDE for the fermion propagator. This occurs because 
perturbative diagrams are dominated in the infrared limit by the lowest available Matsubara frequency, which is zero in bosonic sums. Thus, even though QED3 is infrared finite at zero temperature, problems may arise again at nonzero temperature. This issue has engendered some confusion in the literature. 
Some authors  have noted that an infrared divergence exists, but have ignored it\cite{ignore}, or imposed an infrared cutoff\cite{cutoff}, or assumed that higher order corrections remove the divergence\cite{higher-order}. We have previously shown that, in fact, the infrared divergence is endemic to the theory, but that it does not affect observables if a gauge invariant computation is made\cite{LS2}.

In summary, this paper attempts to utilize the finite temperature Schwinger-Dyson equations to compute the phase diagram for spontaneous chiral symmetry breaking in QED3. For the most part, the computation will be made in the rainbow-ladder approximation in Landau gauge, but will include, for the first time, the full effect of vacuum polarisation. We also address, again for the first time, the infrared divergence problem. The result is a phase diagram for QED3 that we believe is reasonably robust and is numerically quite different from that obtained with the instantaneous or quenched approximations.

\section{QED3 and the Schwinger-Dyson Equations}

We begin with a review of the zero temperature formalism to set notation and to establish zero temperature limits for the subsequent finite temperature numerical work. We also discuss renormalisation in a non-gauge-invariant framework and how this differs from a prescription that is often employed.

\subsection{Zero Temperature Equations and Renormalisation}
\label{ZT-sect}

We only consider fermions in the four-dimensional representation of the Clifford algebra and thus do not introduce the photon Chern-Simons term. Because truncating SDEs in a gauge-invariant manner is an unresolved problem at present, we will include a photon mass term in the lagrangian (written in BPH form):

\be
{\cal L} = -\frac{Z_A}{4} F^2 + Z_F \bar \psi(i \rlap{/}\partial + e_0 {Z_A}^{1/2}\rlap{/}A)\psi  + 
\frac{m_0^2}{2} Z_A A^2  - \frac{1}{2 \xi^2}(\partial\cdot A)^2.
\label{L-eqn}
\ee
Recall that the presence of the photon mass term does not jeopardise renormalisability or masslessness of the physical photon, but is required to restore gauge invariance if symmetry breaking regulators are employed.

The vacuum polarisation tensor can be parameterised in terms of scalar form factors as
\be
\Pi_{\mu\nu}(p) = g_{\mu\nu}\Pi_1(p) + P_{\mu\nu} \Pi(p)
\label{pi-eqn}
\ee
where 
\be
P_{\mu\nu} = g_{\mu\nu} - \frac{p_\mu p_\nu}{p^2}
\ee
is the transverse projection tensor. This expression then leads to the following exact form for the (Landau gauge) photon propagator

\be
D_{\mu\nu}(p) = \frac{-iP_{\mu\nu}}{Z_A p^2-Z_A m_0^2 - \Pi(p) - \Pi_1(p)}. 
\label{D-eqn}
\ee
Finally, the fermion propagator is written in terms of wavefunction and mass scalar functions as

\be
S(p) = \frac{i}{A\rlap{/}p - B}.
\label{S-eqn}
\ee

\begin{figure}[ht]
\includegraphics[width=10cm,angle=0]{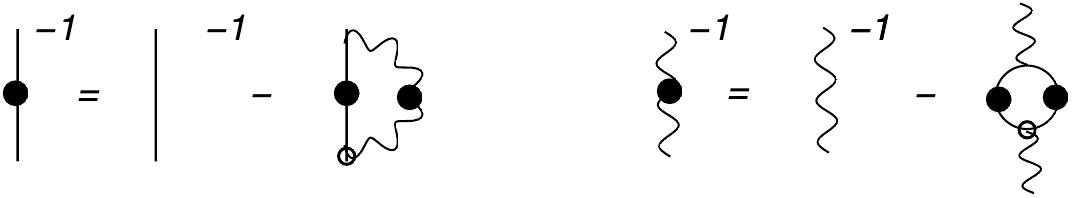}
\caption{Schwinger-Dyson Equations. Solid circles represent full propagators. The open circles represent model vertices.}
\label{sde-fig}
\end{figure}

The scalar functions are obtained by solving the SDEs illustrated in Fig. \ref{sde-fig}. As is well known, these equations couple to higher $n$-point functions and thus must be truncated in some way to yield a tractable problem. A traditional truncation involves using model vertex functions, which we use here. Most of the following results will be in the rainbow-ladder truncation where the vertices are taken to be their bare counterparts. A more sophisticated vertex model, called the Ball-Chiu vertex, obtaining minimal gauge invariance (ie, it is necessary but not sufficient) is written as

\be
i \Gamma^\mu_{\rm BC}(k,p) = \bar A \gamma^\mu + \Delta A \, (k+p)^\mu (\rlap{/}k + \rlap{/}p) - \Delta B \, (k+p)^\mu
\label{BC-eqn}
\ee
where
\be
\bar A = \frac{1}{2}( A(k) + A(p)),
\ee

\be
\Delta A = \frac{A(k)-A(p)}{k^2-p^2},
\ee
and

\be
\Delta B = \frac{B(k)-B(p)}{k^2-p^2}.
\ee
We call $i\Gamma^\mu = \bar A \gamma^\mu$ the ``central Ball-Chiu vertex".

Regulating numerical SDEs is somewhat problematic because it is not possible to find a simple translationally and gauge invariant regulator. This implies that the vacuum polarisation will take on the form shown in Eq. \ref{pi-eqn}. Typically this implies (as it does here) that $\Pi_1$ is divergent, while $\Pi$ is finite. It has been common to avoid this divergence in numerical work by simply projecting it away\cite{BP}

\be
\Pi_{BP} = \frac{1}{2}(g_{\mu\nu} - d\hat p_\mu \hat p_\nu) \Pi^{\mu\nu}(p),
\ee
where $d$ is the spacetime dimension. This is called ``Brown-Pennington projection". One then uses the `usual' photon propagator proportional to $(p^2-\Pi_{BP})^{-1}$ in subsequent work. However, it is not necessary to follow this, rather ad hoc, procedure since traditional renormalisation methods can be used to eliminate the divergence. In this case one projects as
\be
\Pi_1 = \hat p_\mu \hat p_\nu \Pi^{\mu\nu}
\ee
and
\be
\Pi = \frac{1}{2}(g_{\mu\nu} - d\hat p_\mu \hat p_\nu) \Pi^{\mu\nu}(p).
\ee
Notice that the Brown-Pennington and formal expressions for $\Pi$ coincide, but the photon propagator does not. Specifically, requiring that the pole of the photon propagator lies at zero momentum gives $Z_A m_0^2 + \Pi_1(0)=0$ and hence

\be
D_{\mu\nu} = \frac{-i P_{\mu\nu}}{Z_Ap^2 - \Pi(p) - (\Pi_1(p)-\Pi_1(0))},
\label{D-renorm-eqn}
\ee
which is {\it not} the same form as the Brown-Pennington method yields.  We will examine the differences between the BP prescription and BPH renormalisation below.

The Schwinger-Dyson equations are obtained by evaluating the diagrams of Fig. \ref{sde-fig} along with the definitions of Eqs. \ref{L-eqn} -- \ref{BC-eqn}. After rotating to Euclidean space and projecting one obtains

\be
B(p) = 2 Z_F^2 Z_A e_0^2 \int \frac{d^3q_E}{(2\pi)^3} \, B(q) \frac{\bar A(p,q)}{A^2q^2+B^2} \frac{1}{Z_A K^2 + \Pi(K) + \Pi_1(K)-\Pi_1(0)}
\ee
where we show the result with the central Ball-Chiu vertex and $K^2 = (p-q)^2$ (recall that all momenta are Euclidean).  Also
 
\be
A(p) = Z_F + \frac{2 Z_F^2 Z_A e_0^2}{p^2} \int \frac{d^3q_E}{(2\pi)^3}\, A(q) \frac{\bar A(p,q)}{A^2q^2+B^2} \, \frac{\hat K \cdot p \hat K \cdot q}{Z_A K^2 + \Pi(K) + \Pi_1(K)-\Pi_1(0)}.
\ee

Since the integrals in these expressions are finite one can set $Z_F^2 Z_A e_0^2 = 4 \pi \alpha$.  It is convenient to choose MOM-type renormalisation conditions in the Euclidean regime; thus we set $A(M) = 1$ and $B(M) = 0$ where $p_E^2 = M^2$ is a Euclidean mass renormalisation scale. Again, the integrals are convergent, so it is possible to send $M \to \infty$. Lastly, $\Pi(p)+\Pi_1(p)-\Pi_1(0)$ is finite so there is no need to set the photon propagator residue and we can simply take $Z_A=1$. The net result is the renormalised zero-temperature gap equations:

\be
B(p) = 8 \pi \alpha \int \frac{d^3q_E}{(2\pi)^3} \, B(q) \frac{\bar A(p,q)}{A^2q^2+B^2} \frac{1}{K^2 + \Pi_{\rm tot}(K)},
\ee

\be
A(p) = 1 + \frac{8 \pi \alpha}{p^2} \int \frac{d^3q_E}{(2\pi)^3}\, A(q) \frac{\bar A(p,q)}{A^2q^2+B^2} \, \frac{\hat K \cdot p \hat K \cdot q}{K^2 + \Pi_{\rm tot}(K)},
\ee
and
\be
\Pi_{\rm tot}(p) \equiv \Pi(p)+\Pi_1(p)-\Pi_1(0) = -16\pi \alpha\int \frac{d^3q_E}{(2\pi)^3} \, \left[ \bar A(q,Q)\, \frac{BB' + AA' \hat p \cdot q \, \hat p \cdot Q}{(A^2q^2+B^2)({A'}^2 Q^2 + {B'}^2)} - {\rm ditto}(p=0) \right].
\ee
In the last expression $Q^\mu = (p+q)^\mu$, $B=B(q)$, and $B' = B(Q)$ (with similar expressions for $A$ and $A'$). For comparison, we quote the Brown-Pennington form of the vacuum polarisation:

\be
\Pi_{BP}(p) = 16\pi \alpha\int \frac{d^3q_E}{(2\pi)^3} \, \bar A(q,Q)\, \frac{AA'( q\cdot Q - 3 \hat p \cdot q \, \hat p \cdot Q)}{(A^2q^2+B^2)({A'}^2 Q^2 + {B'}^2)} .
\ee

In the following all results will be presented in units of $\alpha$.

Although there is no reason for the two expressions for the vacuum polarisation to coincide, surprisingly, they are very similar. Thus, at least in this case, the Brown-Pennington prescription and standard renormalisation give nearly identical results for the propagator scalar functions and the chiral condensate. In the rainbow-ladder approximation the condensate is $\langle \bar \psi \psi\rangle \approx 0.133$, while in the central Ball-Chiu approximation $\langle \bar \psi \psi\rangle \approx 0.104$. 

The quenched approximation, in which vacuum polarisation is ignored, often appears in the literature. We chose to implement this with a constant polarisation, denoted $\zeta^2$. In this case the condensate in the rainbow-ladder approximation is 

\be
\langle \bar \psi \psi \rangle \approx 0.376 - 0.06 \cdot \zeta^2
\label{RL-cond-eqn}
\ee
and with the central Ball-Chiu vertex is
\be
\langle \bar \psi \psi \rangle \approx 0.346 - 0.15\cdot  \zeta^2.
\label{CBC-cond-eqn}
\ee
These equations are accurate for $\zeta^2$ larger than about 0.1. Remarkably, for very small regulator, both results approach the same value, $\langle \bar \psi \psi\rangle \approx 0.365$.

%
%
%

\subsection{Finite Temperature Formalism}

We employ the imaginary time formalism and choose to work covariantly, which necessitates introducing a three-vector, $n^\mu$, that represents the heat bath. Thus the full fermion propagator is
\be
S = \frac{i}{C\rlap{/}n + A \rlap{/}q - B } \equiv \frac{i}{A_0 n\cdot q\,\rlap{/}n - A \vec \gamma\cdot \vec q - B}
\label{ST-eqn}
\ee
Here $\mu$ is the fermion chemical potential,
\be
p^\mu = (i\omega_n + \mu,\vec p)
\ee
where $\omega_n = (2n+1)\pi T$ is a fermionic Matsubara frequency, and $A$, $B$, and $C$ are functions of $p^\mu$.

The presence of the heat bath generalises the structure of the vacuum polarisation tensor because it is possible to construct a new vector that is orthogonal to $p$:

\be
p^\perp_\mu = p_\mu - n_\mu \frac{p^2}{n\cdot p}
\ee
Thus there are two transverse tensors:

\begin{equation}
P^L_{\mu\nu} =  {\hat p}^\perp_\mu {\hat p}^\perp_\nu.
\end{equation}
and

\begin{equation}
P^\perp_{\mu\nu} = P_{\mu\nu}-P^L_{\mu\nu}.
\end{equation}
The longitudinal and transverse tensors are projections and are orthogonal.

With these definitions and a cutoff regulator one has

\be
\Pi_{\mu\nu} = \Pi_1 g_{\mu\nu} + \Pi_L P^L_{\mu\nu} + \Pi_\perp P^\perp_{\mu\nu}.
\ee

Thus 

\be
\Pi_1 = \hat p_\mu \hat p_\nu \, \Pi^{\mu\nu},
\ee

\be
\Pi_\perp = (P^\perp_{\mu\nu}-\hat p_\mu \hat p_\nu) \Pi^{\mu\nu},
\ee
and
\be
\Pi_L = (P^L_{\mu\nu}-\hat p_\mu \hat p_\nu) \Pi^{\mu\nu}.
\ee


Finally, the photon propagator is (Landau gauge again)

\be
iD_{\mu\nu} = \frac{-iP_{\mu\nu}^\perp}{p^2-\Pi_\perp(p)-\Pi_1(p)+\Pi_1(0)} +\frac{-iP_{\mu\nu}^L}{p^2-\Pi_L(p)-\Pi_1(p)+ \Pi_1(0)}.
\ee
We remind the reader that renormalisation for the finite temperature theory is fixed by the zero temperature limit. Thus we have adopted the same renormalisation conventions as went into deriving Eq. \ref{D-renorm-eqn}.

Notice also that the Ball-Chiu vertex is also generalised in the finite temperature case. Enforcing the Ward-Takahashi identity with a propagator parameterised as in Eq. \ref{ST-eqn} implies that the vertex must contain at least four terms:

\be
i \Gamma_\mu(p,q) = \bar A \gamma_\mu + \Delta A (p+q)_\mu (\rlap{/}p + \rlap{/}q) - \Delta B (p+q)_\mu + \Delta C (p+q)_\mu \rlap{/}n.
\ee
See Ref. \cite{strickland} for a more detailed discussion of the generalised Ball-Chiu vertex.

Explicit expressions for the polarisation scalar functions are
($p^2 = \omega^2 + \vec p^2$):

\begin{eqnarray}
\Pi_1 &=& \frac{16 \pi \alpha}{p^2} T \sum_\nu\int\frac{d^2q}{(2\pi)^2}\, \frac{\bar A}{D} \, \Big[ 2(A_0 \tilde\nu\omega +A\vec q\cdot \vec p)(A_0' \omega(\omega+\tilde\nu) + A' \vec Q\cdot \vec p) - p^2\, (A_0\, A_0' \tilde\nu(\omega+\tilde\nu) + \nonumber \\
&& + AA' \vec q\cdot \vec Q) - BB'p^2)\Big],
\end{eqnarray}

\be
\Pi_\perp = -16 \pi \alpha T \sum_\nu\int\frac{d^2q}{(2\pi)^2}\, \frac{\bar A}{D} \, \left[ 
-2AA' \vec q\cdot \vec Q + 2AA' \hat p \cdot \vec q\, \hat p \cdot \vec Q + \frac{2}{p^2} (\omega\tilde\nu A_0 +\vec p\cdot \vec q A)\,(\omega(\omega+\tilde\nu)A_0' + \vec p \cdot \vec Q A')\right],
\ee
and
\be
\Pi_L = -16 \pi \alpha T \sum_\nu\int\frac{d^2q}{(2\pi)^2}\, \frac{\bar A}{D} \, \left[ 
\frac{4}{p^2}(\omega\tilde\nu A_0+\vec p \cdot \vec q A)\,(\omega(\omega+\tilde\nu)A_0' + \vec p\cdot \vec Q A') - 2A_0 A_0' \tilde\nu(\omega+\tilde\nu) - 2AA' \hat p\cdot \vec q\, \hat p\cdot \vec Q\right],
\ee
where $\tilde \nu \equiv \nu - i \mu$ and 
\be
D = (A_0^2 {\tilde\nu}^2 + A^2 \vec q^2 +B^2)\,({A_0'}^2 (\omega+\tilde\nu)^2 + {A'}^2 \vec Q^2 + {B'}^2).
\ee

The behaviour of the electric and magnetic screening masses is crucial to specifying the properties of the theory. These are defined as

\be
m_{\rm el}^2 = \lim_{p\to 0} \Pi_L^{\rm tot}(0,p) \equiv \lim_{p\to 0} (\Pi_L(0,p)-\Pi_1(0,p))
\ee
and
\be
m_{\rm mag}^2 = \lim_{p\to 0} \Pi_\perp^{\rm tot}(0,p) \equiv \lim_{p\to 0} (\Pi_\perp(0,p)- \Pi_1(0,p)).
\ee
As noted above, in practice it is found that $\Pi_1(p)-\Pi_1(0)$ is very nearly zero, we thus ignore this contribution to the photon propagator in the following discussion. In this case it is simple to show that

%
%
%

\be
m_{\rm mag}^2 =0
\ee
for the rainbow-ladder and central Ball-Chiu vertices considered here. In the perturbative limit where $A\to 1$ and $B \to m$ one obtains

\be
m_{\rm el}^2 = 4 \alpha N_f T \big[ 2 \log 2 + \log(\cosh(\frac{m+\mu}{2T}) + \log(\cosh(\frac{m-\mu}{2T}) - \frac{m}{2T}(\tanh(\frac{m+\mu}{2T}) + \tanh(\frac{m-\mu}{2T})) \big]
\label{mel-eqn}
\ee
Furthermore, one can show that 
\be
\lim_{T\to 0} m_{\rm el}^2 = 0
\ee
in the case with general fermion dressing.
Lastly, it is possible to formally show that the perturbative expressions for $\Pi_L$ and $\Pi_\perp$ approach the perturbative zero-temperature result

\be
\Pi_{pert} = 4\alpha p_E^2 \int_0^1 dx \frac{x(1-x)}{\sqrt{m^2 + x(1-x)p_E^2}},
\ee
as desired.

The equations for the fermion dressing functions are obtained as in the zero temperature case and are 

\begin{equation}
B(\omega,\vec p) = m + 4\pi\alpha T \sum_{\nu} \int \frac{d^2q}{(2\pi)^2}\, \bar A(p,q) \frac{B(\nu,\vec q)}{F(\nu,\vec q)} \left[ \frac{1}{K^2 + \Pi_\perp(K)} + \frac{1}{K^2 + \Pi_L(K)}\right],
\label{B-eqn}
\end{equation}

\be
A_0(\omega,\vec p) \omega = \omega - 4\pi\alpha T \sum_{\nu}\int \frac{d^2q}{(2\pi)^2} \frac{\bar A(p,q)}{F(\nu,\vec q)} \left[ \frac{1}{K^2} ( -2 A(\omega-\nu) \vec q \cdot \vec K - A_0 \tilde\nu ((\omega-\nu)^2 - {\vec K}^2)\frac{1}{K^2+\Pi_L} - \frac{A_0 \tilde\nu}{K^2+\Pi_\perp} \right],
\end{equation}

\begin{eqnarray}
A(\omega,\vec p) {\vec p}^2 &=& {\vec p}^2 + 4 \pi \alpha T \sum_{\nu}\int \frac{d^2q}{(2\pi)^2}\, \frac{\bar A(p,q)}{F(\nu,\vec q)} \Big[ \frac{1}{K^2} ( 2A_0 \vec p \cdot \vec K \tilde\nu(\omega-\nu) + A \vec p\cdot \vec q \, K^2 - 2 \hat K\cdot\vec p\, \hat K \cdot \vec q\, (\omega-\nu)^2) \frac{1}{K^2 + \Pi_L(K)} - \nonumber \\
&&  A\,(\vec p\cdot \vec q -2 \hat K\cdot \vec p\,\hat K \cdot \vec q)\frac{1}{K^2+\Pi_\perp(K)}\Big].
\end{eqnarray}
The notation
\be
K^2 = (\omega-\nu)^2 + (\vec p - \vec q)^2
\ee
and
\be
F(\nu,\vec q) = A_0^2 (\nu-i\mu)^2 + A^2 \vec q^2 + B^2
\ee
has been adopted in these expressions.

Examination of Eq. \ref{B-eqn} reveals an important complication in the formalism, namely when $\nu = \omega$ a logarithmic infrared divergence occurs in the integral over $\vec q$. This divergence is regulated by the electric screening mass in the second term of Eq. \ref{B-eqn}; however, the first term is not regulated by an analogous magnetic screening mass and a divergence must necessarily arise. It is possible to show that this divergence is endemic to the theory at all orders\cite{LS2}. However, if a gauge invariant computation is made, the divergence does not affect observables. Since we must deal with truncated Schwinger-Dyson equations this represents a substantial obstacle to obtaining  reliable results. One of our major conclusions will be that it is indeed possible to make robust statements about QED3 in the finite-temperature Schwinger-Dyson formalism.

\subsubsection{Instantaneous Approximation}

As mentioned in the Introduction, the instantaneous approximation is often used to simplify the analysis of field theories at finite temperature and density. We implement this by neglecting the frequency dependence in the dressed photon propagator and the vertex model. This permits the evaluation of the remaining frequency dependence, yielding the following thermodynamic function

\be
\Theta(T,\mu,E) \equiv T\sum_\nu \frac{1}{(\nu-i\mu)^2 + E^2} = \frac{1}{4E} \left( \tanh\frac{E-\mu}{2T} + \tanh\frac{E+\mu}{2T}\right).
\ee

The fermionic  gap equations simplify (we specialise to rainbow-ladder approximation) to:

\be
A_0 = 1,
\ee
\be
B(p) = 4\pi\alpha \int \frac{d^2q}{(2\pi)^2}\, B(q) \left( \frac{1}{\vec K^2+\Pi_L(\vec K)} + \frac{1}{\vec K^2+\Pi_\perp(\vec K)}\right)\,\Theta(T,\mu,\sqrt{q^2A^2+B^2}),
\ee
and

\be
A(p) \, \vec p^2= \vec p^2 + 4\pi\alpha\int \frac{d^2q}{(2\pi)^2}\, A(q) \left( \frac{\vec p\cdot \vec q}{\vec K^2 + \Pi_L(\vec K)} - \frac{\vec p \cdot \vec q - 2\hat K \cdot \vec p\, \hat K \cdot \vec q}{\vec K^2 + \Pi_\perp(\vec K)}\right) \, \Theta(T,\mu,\sqrt{q^2A^2 +B^2}).
\ee
Notice that the dressing functions are no longer functions of Matsubara frequencies.

The efficacy of this approximation will be tested in the next section. For now we remark that the neglect of frequency-dependence implies that it is impossible to recover the low temperature limit.

\section{The QED3 Phase Diagram}

QED3 at finite temperature is a computationally intensive problem because the Matsubara sums do not converge quickly; thus it is equivalent to a large collection (approximately 100) of coupled zero-temperature problems. Perhaps more vexing is that the gap equations and the vacuum polarisation functions are sensitive to cutoff and other numerical choices (such as momentum grids and interpolation methods). 

We have numerically confirmed that the zero temperature limit is recovered by a sufficiently large Matsubara cutoff. We found that it was crucial to use an $O(3)$-invariant cutoff when evaluating frequency and momentum sums if the zero temperature limit was to be recovered:

\be
\int^\Lambda d^3q \to iT \sum_{n=-\Lambda/2\pi T}^{\Lambda/2\pi T} \int^{\sqrt{\Lambda^2-\nu_n}} d^2q.
\ee

Furthermore, one requires $\Lambda \agt 100 T$. We also confirmed that perturbative expressions are recovered and that an $O(3)$ invariant functional dependence is seen at low temperature.

As will be shown, the value of the electric screening mass is particularly important to chiral symmetry breaking. It is also sensitive to numerical truncations and therefore we found it convenient to evaluate the screening mass separately. Because it is not practical to choose large Matsubara cutoffs we found it useful to add the value of the perturbative screening mass (Eq. \ref{mel-eqn}) evaluated outside of the cutoff $\Lambda$ and subtract its value inside the cutoff. This extends the effective integration region to infinity and proved quite accurate because the dressing functions $A0$, $A$, and $B$ approach their perturbative limits rapidly in the Euclidean momentum. As with the screening mass, it was useful to extend the region of integration beyond the cutoff when evaluating the (subtracted) photon dressing functions. This was achieved with the Ansatz

\be
\Pi(\omega,p) \to \frac{\alpha N_f \pi}{2} \sqrt{\omega^2+p^2},
\ee 
which was confirmed to work well. Evaluation of the photon dressing functions is very time-consuming. We therefore precomputed these functions and iterated the fermionic gap equations until convergence was achieved, recomputed the photon dressing functions, and iterated until the global error dropped below a fixed tolerance.

Finally, a useful approximation to the electric screening mass is obtained if one employs Eq. \ref{mel-eqn} with $m = B(0,0)/A(0,0)$. This is especially helpful when computational constraints prevent an accurate estimate of the screening mass (often at low temperatures).

\subsection{Quenched QED3}

\begin{figure}[ht]
\includegraphics[width=8cm,angle=0]{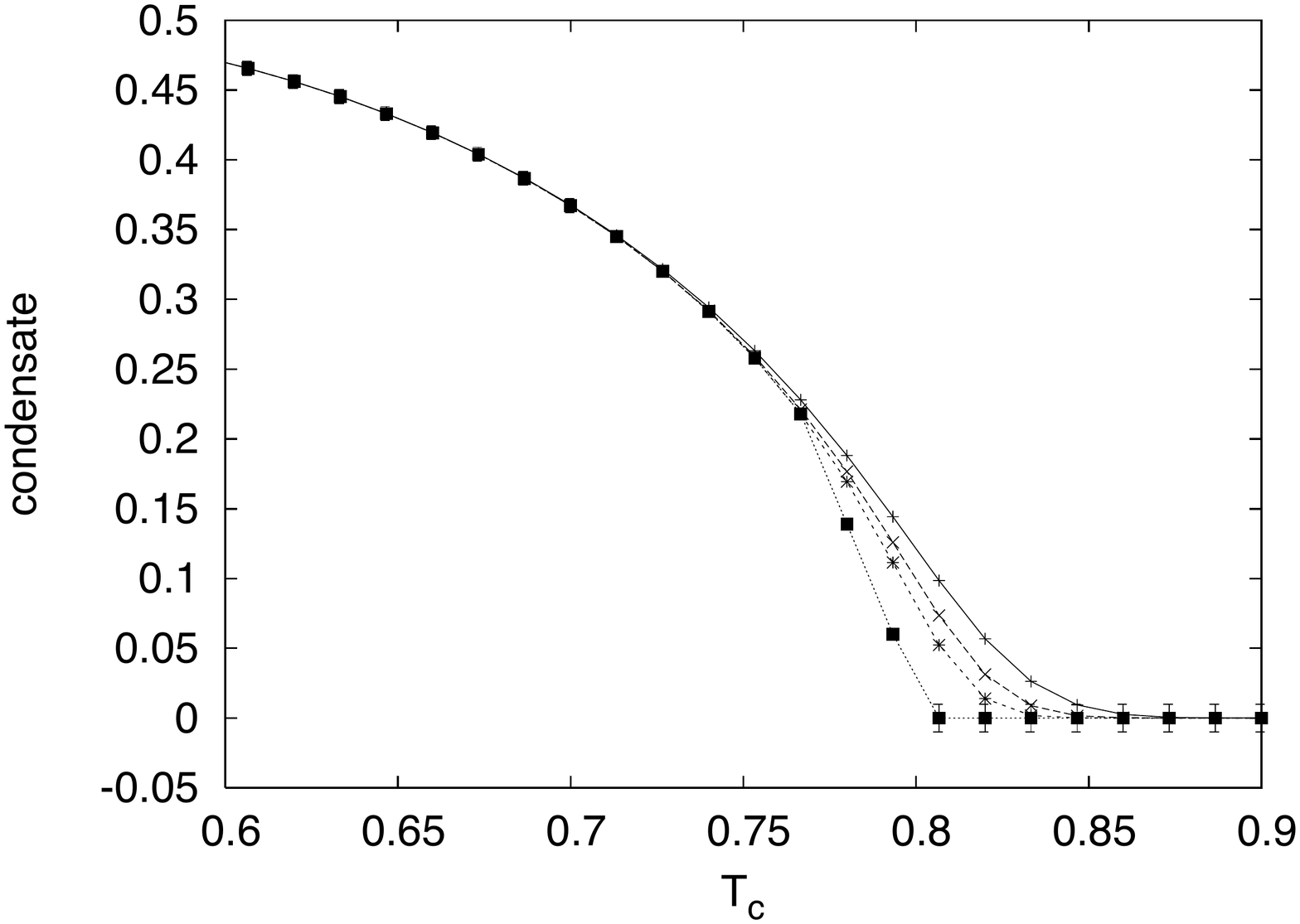}
\qquad\qquad
\includegraphics[width=8cm,angle=0]{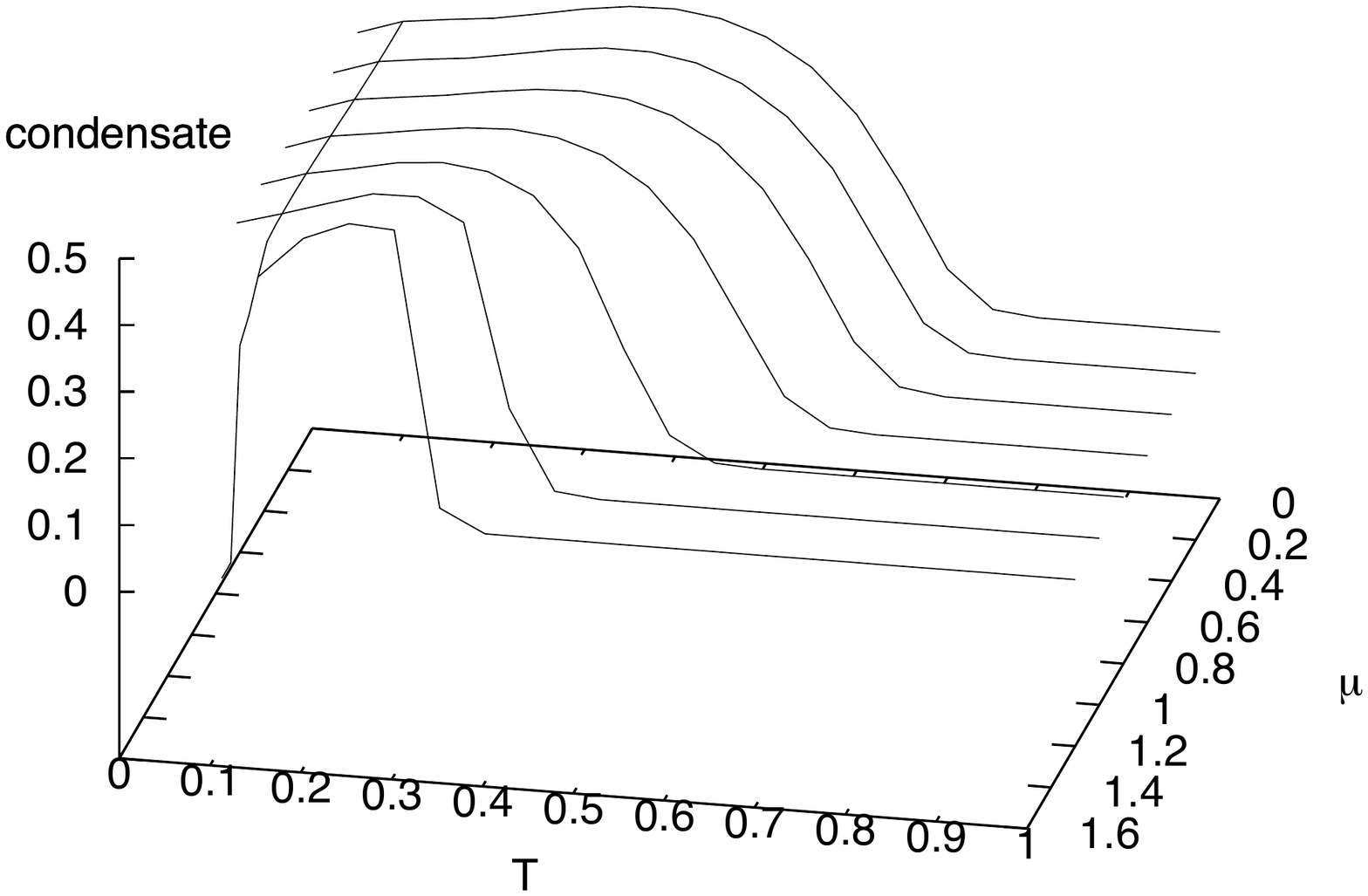}
\caption{(left) Quenched QED3 extrapolation for $\mu=0.4$ and $\zeta_{\rm el}^2=\zeta_{\rm mag}^2 = 0.05$. From top to bottom the curves are for $N_{\rm it}$ = 30, 50, 80, and extrapolated to infinity. (right) Chiral condensate for quenched QED3 with $\zeta_{\rm el}^2 = \zeta_{\rm mag}^2 = 0.05$.}
\label{quenched-fig}
\end{figure}

The simplest numerical case is quenched QED3 wherein we let $\Pi_\perp \to \zeta_{\rm mag}^2$ and $\Pi_L \to \zeta_{\rm el}^2$. Results for the chiral condensate with $\mu = 0.4$ and $\zeta_{\rm el}^2 = \zeta_{\rm mag}^2 = 0.05$ are shown in Fig. \ref{quenched-fig} (left) as a function of temperature (recall that all dimensionful quantities are measured in units of $\alpha$).  One sees an apparent rapid crossover near $T= 0.82$. As is typically the case, convergence is slow near critical points and one must be careful in judging the order of phase transitions. The figure shows additional computations with a larger number of iterations and an extrapolation to the infinite limit. We have found that an approach proportional to $1/\sqrt{N_{it}}$ for $T<T_c$ fit the data quite well. Above the critical temperature the approach was exponential in the number of iterations. It is evident that a second order chiral restoration phase transition is occurring at $T_c \approx 0.8$. We have also confirmed that the zero temperature condensates of Eqs. \ref{RL-cond-eqn} and \ref{CBC-cond-eqn} are reproduced at better than the one  percent level.

The phase diagram for quenched QED3 with $\zeta_{\rm mag}^2 = \zeta_{\rm el}^2 = 0.05$ is shown in Fig. \ref{quenched-fig} (right). The transition is second order everywhere but becomes sharper as the chemical potential rises. The region is bounded by $T_c(\mu=0) \approx 0.89$ and $\mu_c(T=0) \approx 1.4$.

\begin{figure}[ht]
\includegraphics[width=10cm,angle=0]{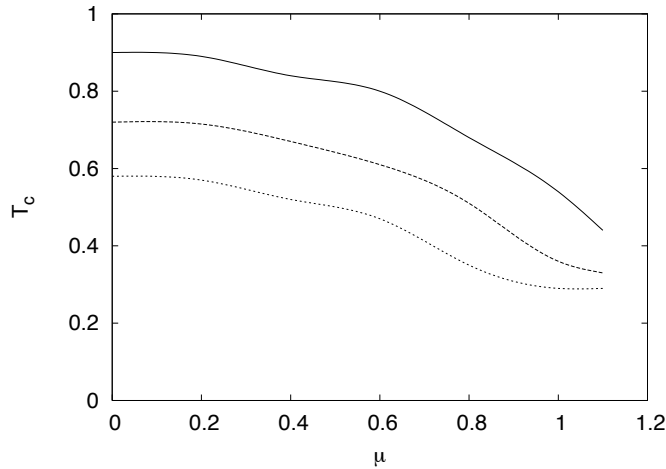}
\caption{Quenched phase diagram. From top to bottom the curves correspond to $\zeta_{\rm el}^2 = \zeta_{\rm mag}^2 = $ 0.05, 0.1, 0.2.}
\label{quenched2-fig}
\end{figure}

Recall that an infrared divergence is exposed as $\zeta_{\rm mag}\to 0$ (since this is the quenched case, $\zeta_{\rm el} \to 0$ is also problematic). We have argued that the divergence does not affect observables, such as the chiral restoration temperature, if a gauge invariant truncation is made\cite{LS2}. However, the quenched approximation does not respect gauge invariance and one expects significant infrared cutoff dependence in the phase diagram. That this is indeed the case is shown in Fig. \ref{quenched2-fig}. where the variation in the phase boundary is shown for $\zeta_{\rm el}^2 = \zeta_{\rm mag}^2 =$ 0.05, 0.1, and 0.2. 

\subsection{Instantaneous QED3}

The instantaneous approximation is often justified by noting that it represents the leading infrared behaviour of the theory. However, this does not guarantee that it is numerically accurate. Accuracy can be simply checked in the quenched case by comparing to the full formalism. We have found a drastic difference from the results of the preceding section. For example, the condensate for $\zeta^2 = 0.05$ and for low temperature is $\langle\bar \psi \psi\rangle \approx 4.5$, which should be compared to the full quenched results of $\langle \bar \psi \psi \rangle \approx 0.365$ (rainbow-ladder approximation was used in both cases). The critical temperature in this case was determined to be $T_c(\mu=0) \approx 2.6$ (to be compared with 0.89 in the full case) and $\mu_c(T=0) \approx 3.3$ (1.4 in the full case). 

It is thus clear that the instantaneous approximation, while a useful computational tool, is numerically unreliable. Perhaps this should not be surprising, the instantaneous approximation is useful in the nonrelativistic weak binding limit where the large scale separation between the fermion kinetic energy and the photon energy permits integrating out the photon degrees of freedom and leaves a potential interaction. But we are concerned with spontaneous chiral symmetry breaking and massless fermions -- a situation far removed from that being considered in the instantaneous approximation.

\subsection{Screened QED3}

We now turn to the case of screened QED3, where the fermion is permitted to feed back into the photon propagator. In this case the electric screening mass prevents divergences in the perpendicular portion of the gap equations. We thus set $\zeta_{\rm el}=0$ in the following. The magnetic portion requires infrared regulation and we retain $\zeta_{\rm mag} >0$. In a gauge invariant computation observables such as the transition temperature would not depend on the value of $\zeta_{\rm mag}$. The sensitivity of the transition temperature to the cutoff then serves as a useful diagnostic for the efficacy of the truncations employed in this work.

\begin{figure}[ht]
\includegraphics[width=10cm,angle=0]{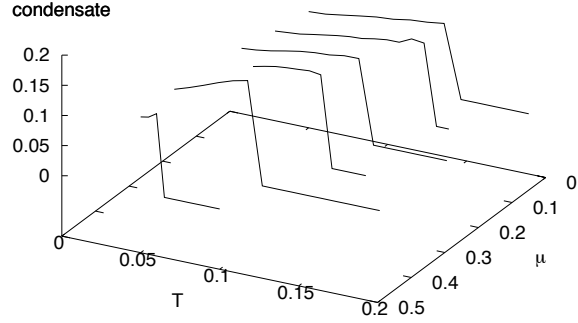}
\caption{Chiral condensate vs. temperature and density ($\zeta^2_{\rm mag} = 0.05$).}
\label{full-phase-fig}
\end{figure}

The condensate is shown as a function of temperature and chemical potential in Fig. \ref{full-phase-fig} for $\zeta_{\rm mag}^2 = 0.05$ for the rainbow-ladder case. One observes a much-reduced condensate, as is expected from the zero temperature results of section \ref{ZT-sect}. Furthermore the region of chiral symmetry breaking is substantially smaller than that of the comparable quenched case. We find $T_c(\mu=0) \approx 0.14$ and $\mu_c(T=0) \approx 0.6$. Although it is difficult to be definitive, it also appears that the phase transition is now first order.

\begin{figure}[ht]
\includegraphics[width=8cm,angle=0]{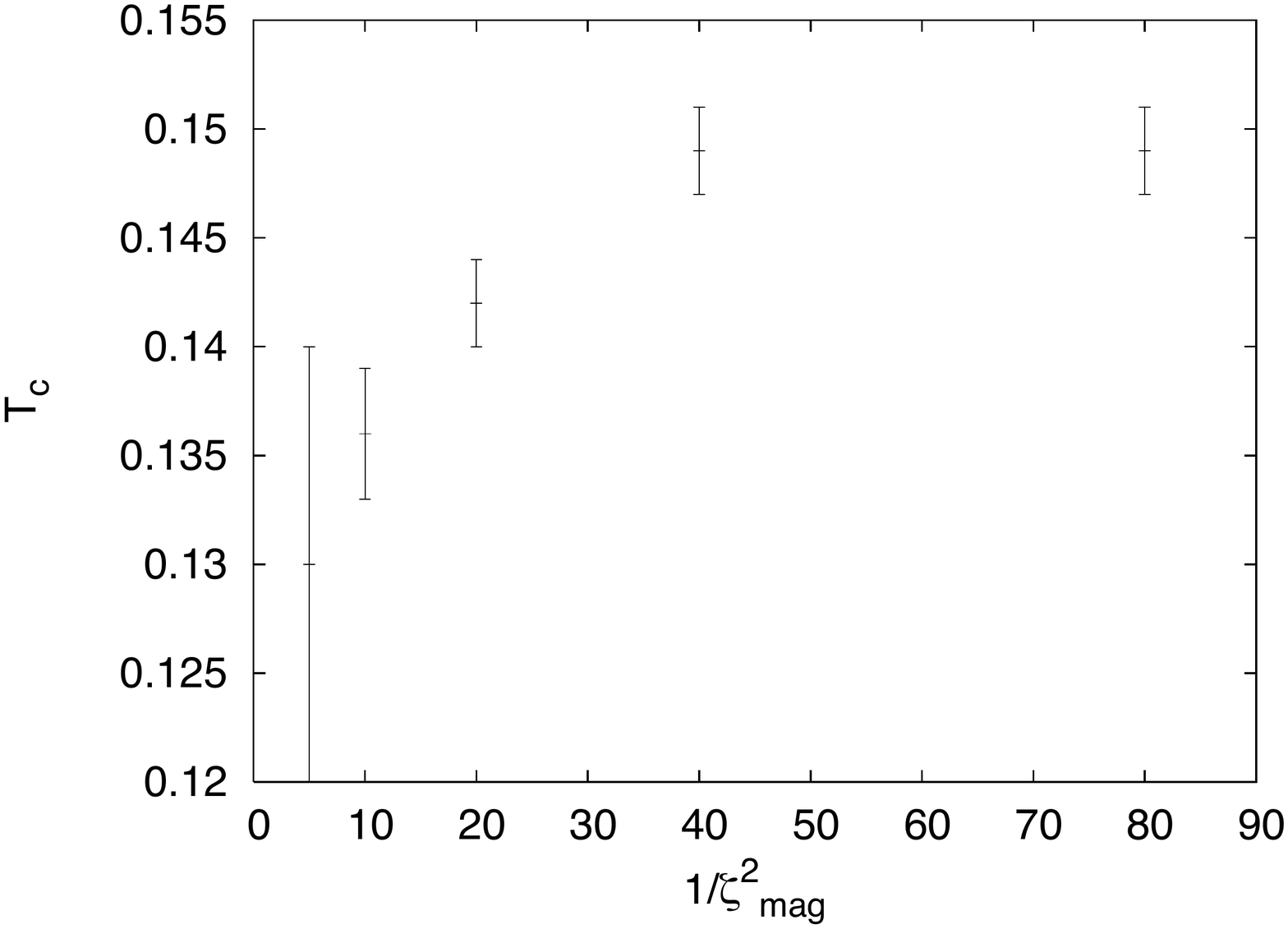}
\qquad\qquad
\includegraphics[width=8cm,angle=0]{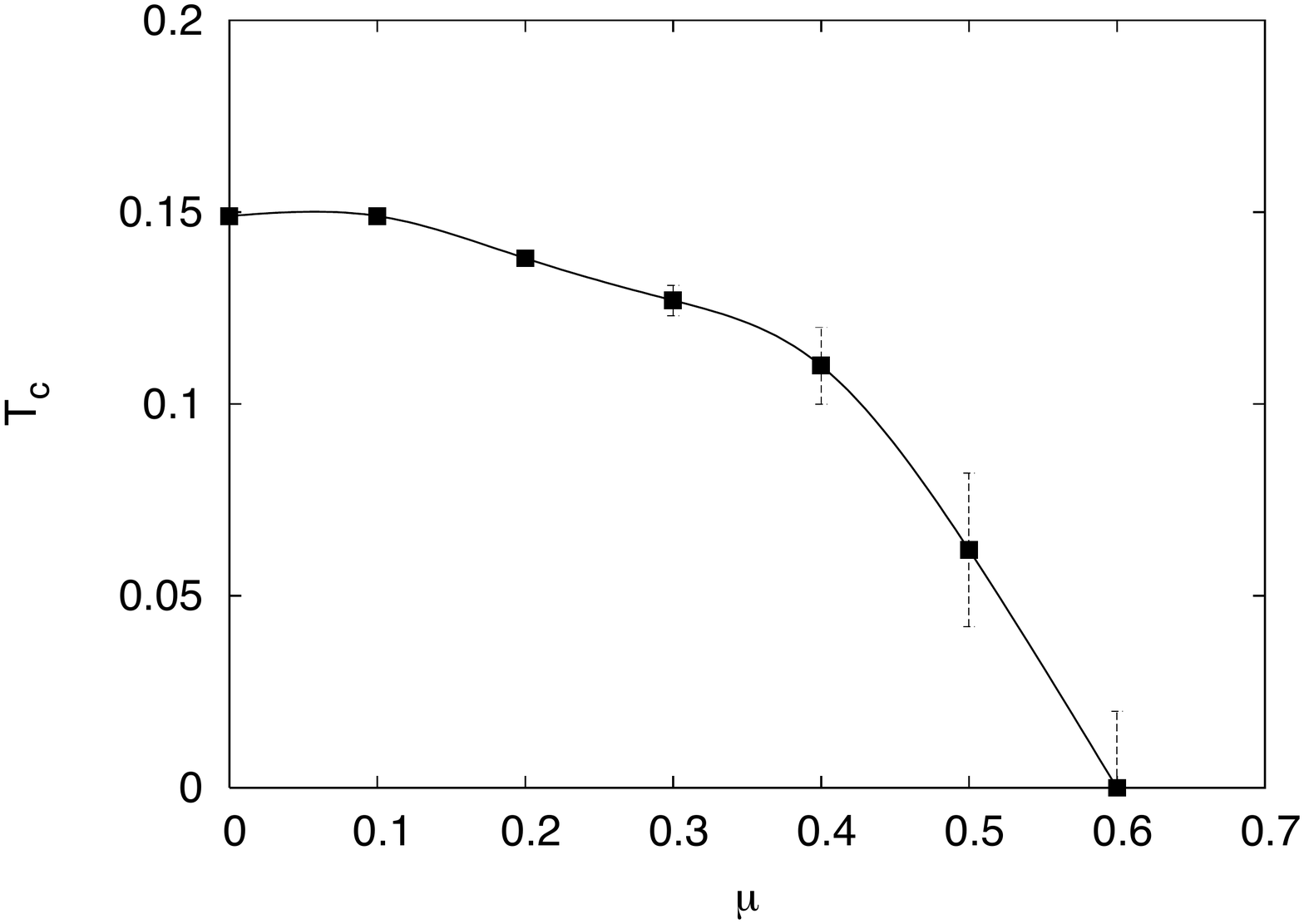}
\caption{Critical Temperature for $\mu=0$ vs. $\zeta^{-2}_{\rm mag}$ (left). Phase diagram for QED3 with vacuum polarization (right).}
\label{full-fig}
\end{figure}

The central question is whether this result is stable under variations in the infrared cutoff, $\zeta_{\rm mag}$. The dependence of the critical temperature on the cutoff is shown in Fig. \ref{full-fig}. As can be seen the critical temperature appears to be approaching a stable value as the cutoff is removed. In contrast, the value of the condensate depends on the cutoff. For example

\be
\langle \bar \psi \psi\rangle(T=0.05,\mu=0) \approx \frac{0.058}{(\zeta_{\rm mag}^2)^{0.153}}.
\ee

Because of the stability of the critical temperature with respect to removing the infrared cutoff, it is possible to construct a reliable phase diagram, shown in Fig. \ref{full-fig} (right). The error bars in this figure incorporate uncertainty in extrapolating to the critical temperature and in sending $\zeta_{\rm mag}^2\to 0$.

\begin{figure}[ht]
\includegraphics[width=10cm]{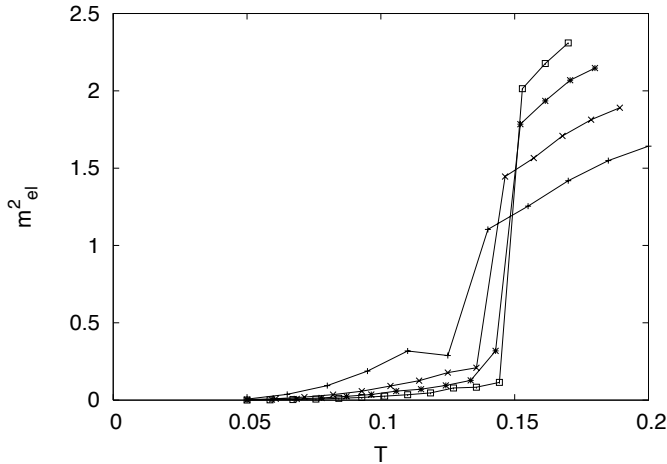}
\caption{Electric Screening Mass vs. temperature for $\mu=0$ (rainbow ladder). Squares: $\zeta_{\rm mag}^2 = 0.0125$, bursts: $\zeta_{\rm mag}^2 = 0.025$, crosses: $\zeta_{\rm mag}^2 = 0.05$, plusses: $\zeta_{\rm mag}^2 = 0.1$.}
\label{mel-fig}
\end{figure}

The sudden transition seen in Fig. \ref{full-phase-fig} is strongly correlated with the electric screening mass, which is very small below the critical temperature and jumps to a large value above it (see Fig. \ref{mel-fig}). It is evident that the transition sharpens rapidly as the infrared cutoff is reduced and it is likely that $m_{\rm el}^2$ undergoes a phase transition itself. Since the polarisation tensor is gauge invariant for abelian gauge theories, the screening mass should be independent of the cutoff as $\zeta_{\rm mag} \to 0$. While it is possible that the screening mass is stabilising above the transition temperature, we judge that the rather large cutoff dependence evident in the figure is a reflection of the truncations made in this study.

\section{Discussion and Conclusions}

To our knowledge, this work represents the first computation of the fermion and gauge boson dressing functions with full frequency-dependence in the Schwinger-Dyson equations. Furthermore, it is the first computation that seriously addresses the infrared divergence that must appear in QED in three dimensions. We have shown that, even in rainbow-ladder approximation, the residual gauge-dependence is not sufficient to invalidate the zero infrared cutoff limit. Thus a reasonably reliable phase diagram is obtained. Of course, more accurate vertex models or extending the computation to higher $n$-point functions is required to assess the accuracy of our result.

We have also determined that the quenched approximation to the photon propagator drastically changes the characteristics of the theory. The condensate and fermion mass function are much larger, the chiral symmetry restoration transition becomes second order, and of course, the observables are sensitive to the infrared cutoff.

Similarly, the oft-used instantaneous approximation has been shown to be inaccurate. Again, this is not surprising since this approximation is only appropriate in the weak coupling, heavy fermion limit.

Finally, we have examined the issue of gauge-noninvariance induced by using a cutoff regulator. Simply projecting onto the transverse structure of the vacuum polarisation tensor has been suggested, and is often used, as a resolution of this problem. We have shown that this is not necessary, and that following the standard renormalisation procedure (with suitably generalised lagrangian) is sufficient to avoid any problems.

\acknowledgments
This research was supported by the U.S. Department of Energy under contract
DE-FG02-00ER41135 (Swanson) and the Frankfurt Institute for Advanced Studies (Lo).

\end{document}